# The influence of building interactions on seismic and elastic surface waves

B. Ungureanu[1], S. Guenneau[2], Y. Achaoui[3], A. Diatta[2],
M. Farhat[4], H. Hutridurga[5], R.V. Craster[1], S. Enoch[2], S. Brûlé[2]

[1] Imperial College London, Department of Mathematics, SW7 2AZ, United Kingdom
[2] Aix Marseille Univ, CNRS, Centrale Marseille, Institut Fresnel, 13013 Marseille, France
[3] FEMTO-ST, CNRS, Université de Bourgogne Franche-Comté, 25044 Besançon, France
[4] Division of Computer, Electrical, and Mathematical Sciences and Engineering, King Abdullah University of Science and Technology (KAUST), Thuwal 23955-69100, Saudi Arabia
[5] Indian Institute of Technology Bombay, Department of Mathematics, SW7 2AZ, India

b.ungureanu@imperial.ac.uk

*Abstract* – **We outline some recent research advances on the control of elastic waves in thin and thick plates, that have occurred since the large scale experiment [*Phys. Rev. Lett.* 112, 133901, 2014] that demonstrated significant interaction of surface seismic waves with holes structuring sedimentary soils at the meter scale. We further investigate the seismic wave trajectories in soils structured with buildings. A significant substitution of soils by inclusions, acting as foundations, raises the question of the effective dynamic properties of these structured soils. Buildings, in the case of perfect elastic conditions for both soil and buildings, are shown to interact and strongly influence elastic surface waves; such site-city seismic interactions were pointed out in [*Bulletin of Seismological Society of America* 92, 794-811, 2002], and we investigate a variety of scenarios to illustrate the variety of behaviours possible.**

## 1. Introduction

We assume that smart cities of the future will increasingly include more buried structures, especially in regions with large variations of temperature with the motivation being, for instance, to improve energy savings. This important substitution of soil by rigid or void elements will inevitably raise the question of the real dynamic response of the sedimentary soils at the free surface. In this article we explore what the theoretical tools borrowed from transformation optics can bring to the design of structured soils such as for a large scale photonic elastic crystal experimentally tested in [1], and we envision what a metacity cloak might achieve. Allied to the discussion of soil modification is the dynamic influence of buildings upon the elastic waves and upon each other.

A

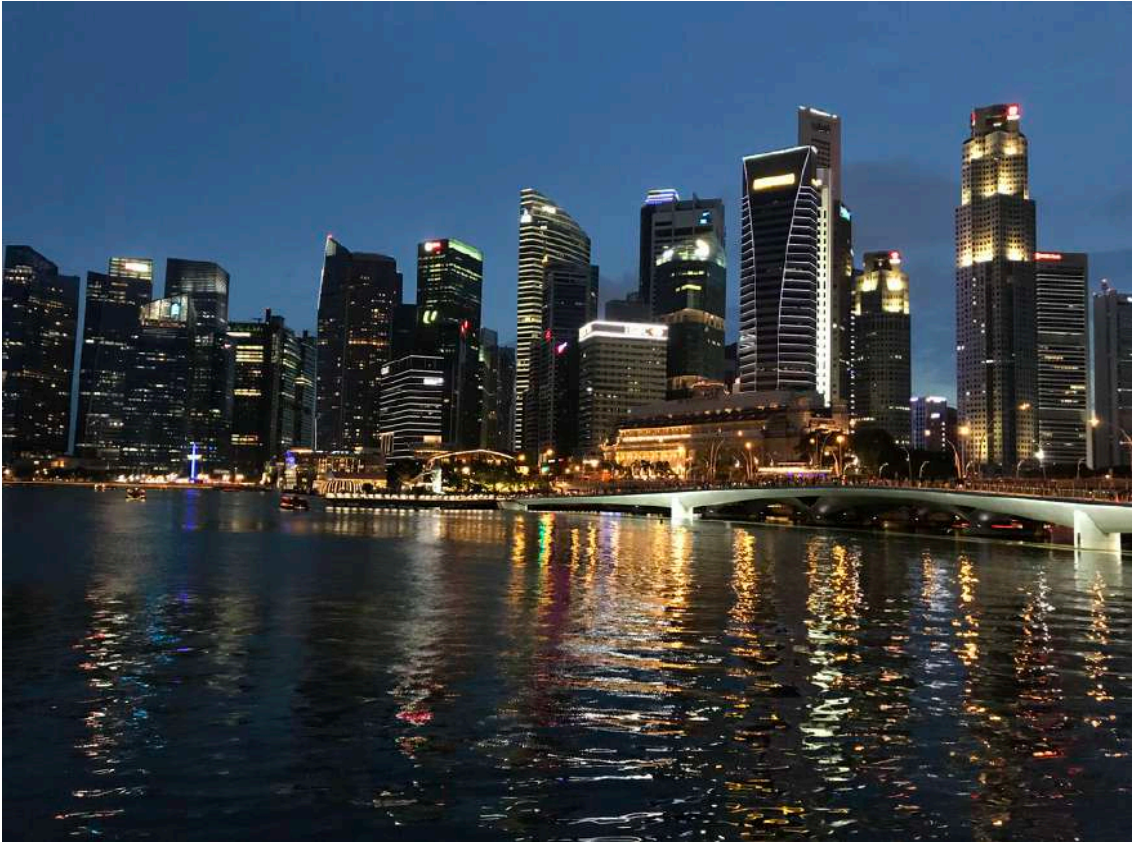

B

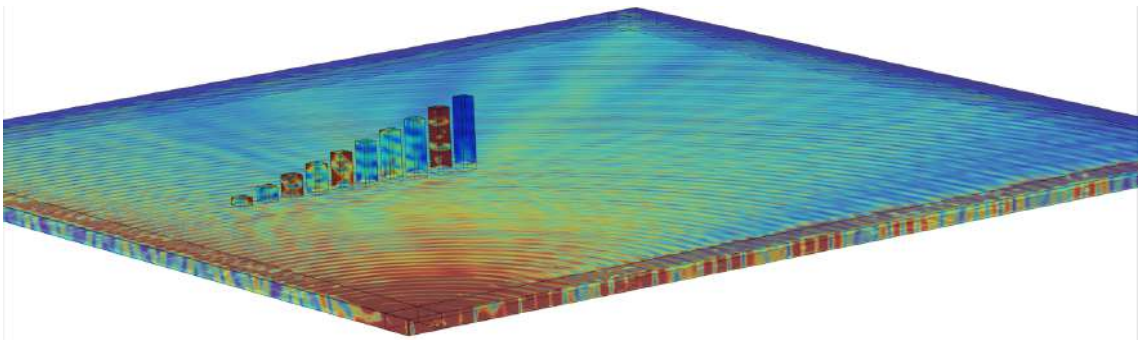

Fig. 1. Photo (A) of the waterfront in Singapore (courtesy of B. Ungureanu) and COMSOL simulation (B) of a seismic wave (of typical frequency 50 Hz) interacting with a cluster of buildings 20m x 20m in cross-section of increasing heights (10 m to 100 m) with centre to centre spacing of 40 m along the propagation path, this is reminiscent of a meta-wedge [2]. In light of the seminal work by Guéguen et al. [3], we know that the high storey buildings in the background can be viewed as a metacity which could display metamaterial-like locally resonant features when a seismic wave propagates through it [4]. An elastic rainbow [5] effect could then occur, which would filter the seismic wave frequencies.

Seventeen years ago, a group of research scientists [3] made the bold claim that a megalopolis such as Mexico-city could have tremendous interactions with seismic waves, and deeply impact the registered ambient seismic noise. At that time, this research grouping might not have been fully aware of the deep consequences their work would have a decade afterwards, following the rise of metamaterials [6] for wave control. We show in figure 1(A) the water front of Singapore, with its skyline, and for a physicist well versed in locally resonant structures such as metamaterials, high-storey buildings that can conceivably couple to waves with a wavelength much larger than their cross-sectional size. Actually, one can view such a cluster of buildings as a scaled-up forest of 'concrete trees'. Some plate experiments with forests of metallic rods performed a few years ago by the group of Roux at University of Grenoble, have indeed shown significant filtering effects for Lamb waves [7], and these experiments motivated experiments on filtering effects of Rayleigh waves in a forest of trees [8].

Then, with the viewpoint that building are potential resonators it becomes clear that the Rayleigh waves could hybridize with the buildings [9] and behave in a way conceptually similar to spoof plasmon polaritons in optics. A collection of buildings such as that at the waterfront of Singapore can be idealised using an elastic rainbow [2], as one can note that there is a gradient in the height of buildings in figure 1(A), and COMSOL simulations then predict so-called rainbow trapping in figure 1(B), where some of the buildings can be seen to resonate with the seismic signal and other ones do not. This is reminiscent of the experimental and numerical analysis of the impact of three closely located towers in the city of Grenoble on local seismicity, and its perception by habitants, during an earthquake of small magnitude [10]. This study supports the hypothesis that buildings in urban areas need not, or indeed should not, be considered as stand-alone constructions with perfect embedding, is usually studied as being independent dynamically. Our simulation in figure 1(B) of an elastic rainbow effect has been investigated previously for forest of trees of an increasing height [2], and in the present case it could prove uncomfortable dangerous for inhabitants of the resonant buildings.

It is even possible that such a large scale meta-wedge [2] could convert some surface seismic wave activity into downward propagating waves, and thus shield other buildings within the megalopolis. The concept of the elastic rainbow is borrowed from the optics community [5], and this suggests intimate links between waves in optics and geophysics. For instance, during the summer of 2012, the civil engineering team of Stéphane Brûlé at the Menard company (Vinci Group) performed the first large scale experiments on a seismic shield [1] near the city of Grenoble, making use of some analogies with photonic crystals within which light is disallowed to propagate over certain frequency ranges, and the same team subsequently tested a flat seismic lens via negative refraction [11] near the city of Lyon. These two experiments confirmed that concepts from the physics of photonic crystals can be translated to the control of Rayleigh waves in structured sedimentary soils. Actually, analogies between optics and geophysics have been further extended to encompass locally resonant structures such as metamaterials going beyond the control of electromagnetic waves thanks to low frequency stop bands [6]. For instance, inertial resonators buried in the soil have been proposed to achieve shielding effects for low frequency seismic waves [12].

It will not therefore be a surprise that the theory of transformation optics introduced by Pendry, Schurig and Smith [13] and conformal optics introduced by Leonhardt [14], has had such a tremendous impact on the elastic wave community in the past ten years. Following Milton, Willis and Briane [15], transformed elastodynamic equations have been studied for bulk elastic waves [16-18] and surface elastic waves in thin plates [19-22].

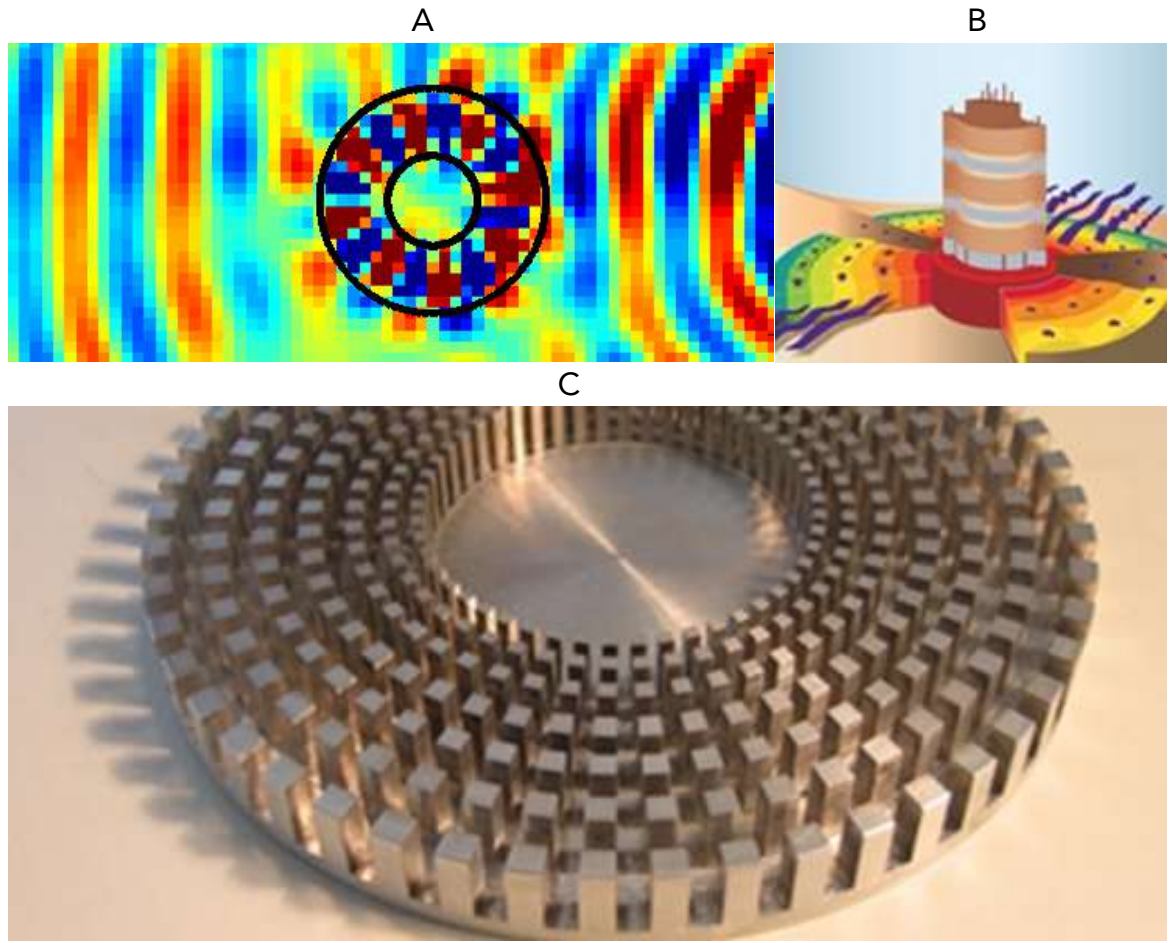

Fig. 2.Cloaking and wave analogies: Water waves, [23] and also microwaves and sound waves [24], experiments have all been performed with the same geometrical cloak. The metamaterial cloak works over a finite interval of wave wavelengths that do not exceed the cloak's diameter (20 cm) and for which the cross-section of its structural elements can be considered small enough (not exceeding one third of the wavelength); (A) microwave experiment (courtesy of R. Abdeddaim and E. Georget, Institut Fresnel) showing both reduced field in the centre of the cloak and cloak's transparency for a transverse electric wave at 3.5 GHz [21]; (B) artistic view of a seismic cloak (Institut Fresnel/INSIS/CNRS); (C) photo of the fabricated aluminium cloak (courtesy of S. Enoch).

We show in figure 2 the power of some wave analogies drawn between optics, hydrodynamics [23,24] and geophysics, that led to the concept of an invisibility cloak (figure 2(C)), working for microwaves (figure 2(A)), and that can serve as a basis for a design of a seismic cloak (figure 2(B)). The ideal goal of a physicist might be to render buildings invisible to seismic waves, but it is not a simple task, as civil engineers know. The design of a mechanical cloak as experimentally achieved by the group of Wegener [25] remains a challenge in the dynamic regime, so controlling in-plane elastic wave trajectories is still an open problem. Of course, some great progress has been made on the control of out-of-plane, flexural, and Rayleigh-like, wave trajectories in thin and thick plates [19-32], but there are other sorts of deleterious, surface Love, and bulk, seismic waves, that cannot be satisfactorily controlled yet, at least from a theoretical stand-point [33]. One might argue that at least controlling Rayleigh waves [34] would greatly reduce the damaging effect of an earthquake in sedimentary soils, as any buildings located inside the seismic zone would be in theory untouched by Rayleigh waves. There are other issues that need thought, for instance, there is a deep connection between cloaking and the theory of inverse problems, in the context of partial differential equations [35], that suggests a seismic cloak might just have the opposite effect as what one wishes to achieve. Indeed, from the standpoint of inverse problems [35], a perfect cloak should make the reconstruction of material properties based on boundary measurements impossible. Intuitively, an aspect of cloaking is that waves cannot penetrate the invisibility region within the cloak, so one expects that this can have interesting applications in wave protection. However, some fundamental theorems in inverse problems [36], show that while one could not reconstruct material properties within the cloak and its core, some trapped modes might occur (associated with properties of the so-called Dirichlet to Neumann map). Such trapped modes in the context of a seismic cloak might be even more devastating than the incoming seismic wave would have been in the absence of cloak! We will show a range of these features in the next section.

## 2. Transformation seismology for the design of metacity cloaks

We now investigate the possibility of using interacting buildings and waves to try and create a design of a metacity that could have the ability to self-protect itself or, at least, the centre of the city against earthquakes. We proceed as follows: first we consider a geometry like in figure 2(C) with concentric layers of pillars. From our knowledge of transformation optics, we know that such a metamaterial corresponds to transformed anisotropic parameters that create an exclusion zone inside which an object is made invisible (known as a hole in electromagnetic space). Such a cloak, designed for electromagnetic waves, tells us that a scaled-up cloak of increasing anisotropy towards its centre might well behave in a similar manner for elastic waves, see e.g. for a seismic cloak designed with Maxwell-fisheyes or Luneburg lenses [37]: an exclusion zone could be created and a building placed inside this zone would be protected from elastic waves. Then we approximate the elastic parameters of these layers using an engineering algorithm that amounts to placing graded buildings of increasing height when one moves from the outer concentric ring of the cloak towards the inner ring, see figure 3. Different designs have been tested and our aim here is to show the range of possible behaviours.

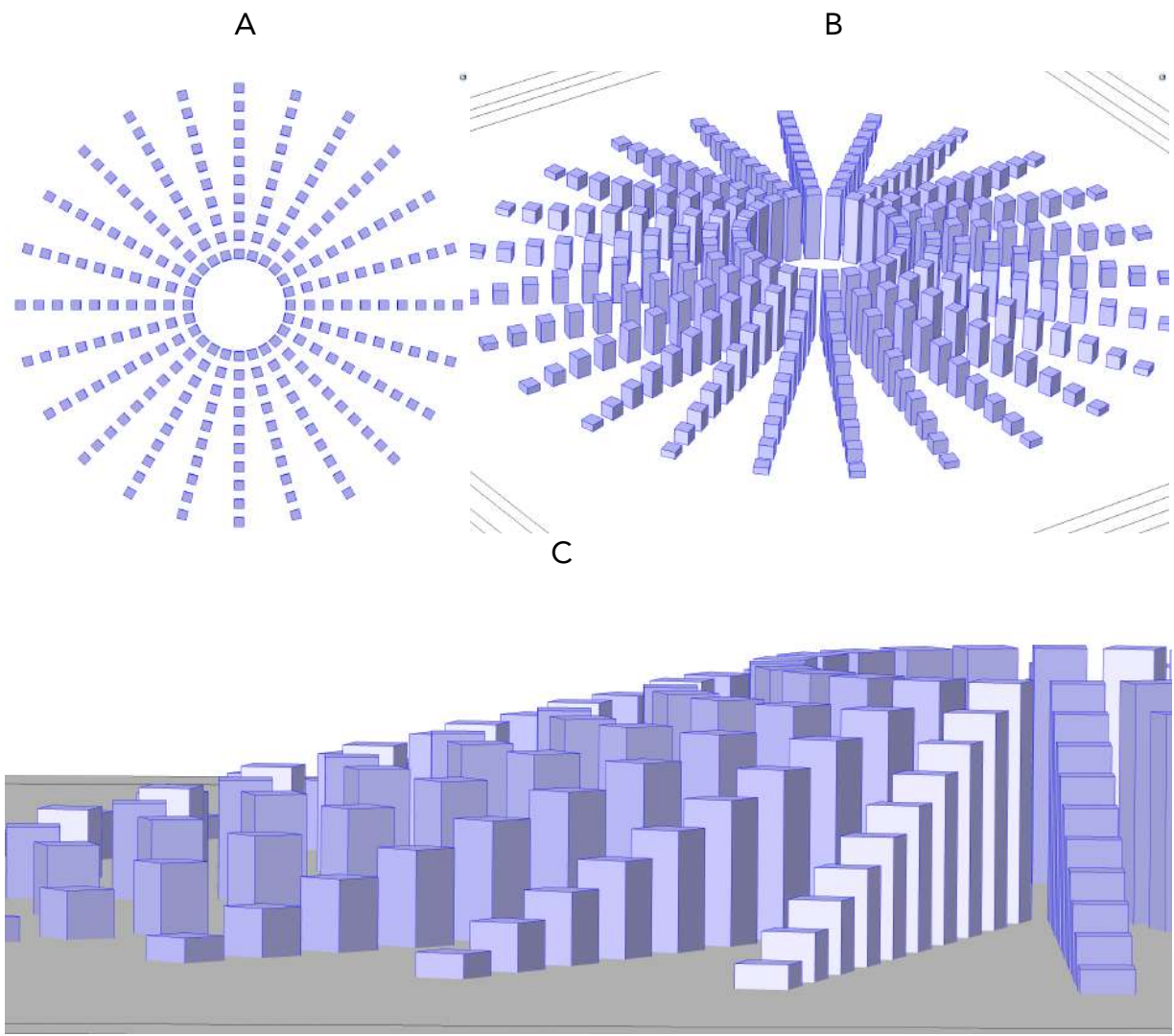

Fig. 3. Design of a large scale elastodynamic cloak in a plate with 10 concentric layers of pillars simulating buildings with height varying linearly from 10 m to 100 m, of square cross section 20 m x 20 m, with centre to centre spacing of 40m along the propagation path. The plate has thickness of 40 m and the cloak has diameter of 1000 m, with a protected area in the centre which is 200 m in diameter and buildings have parameters of concrete ($E$ = 25e9 [Pa], $\rho$ = 2300 [kg/m$^3$], $v$ = 0.2) with 20 m of interstitial distance in between.

Such a plate design should work for Rayleigh waves, but not for in-plane elastic waves such as surface Love waves or coupled shear and pressure elastic waves, which would require some twisting modes (elastic chirality).

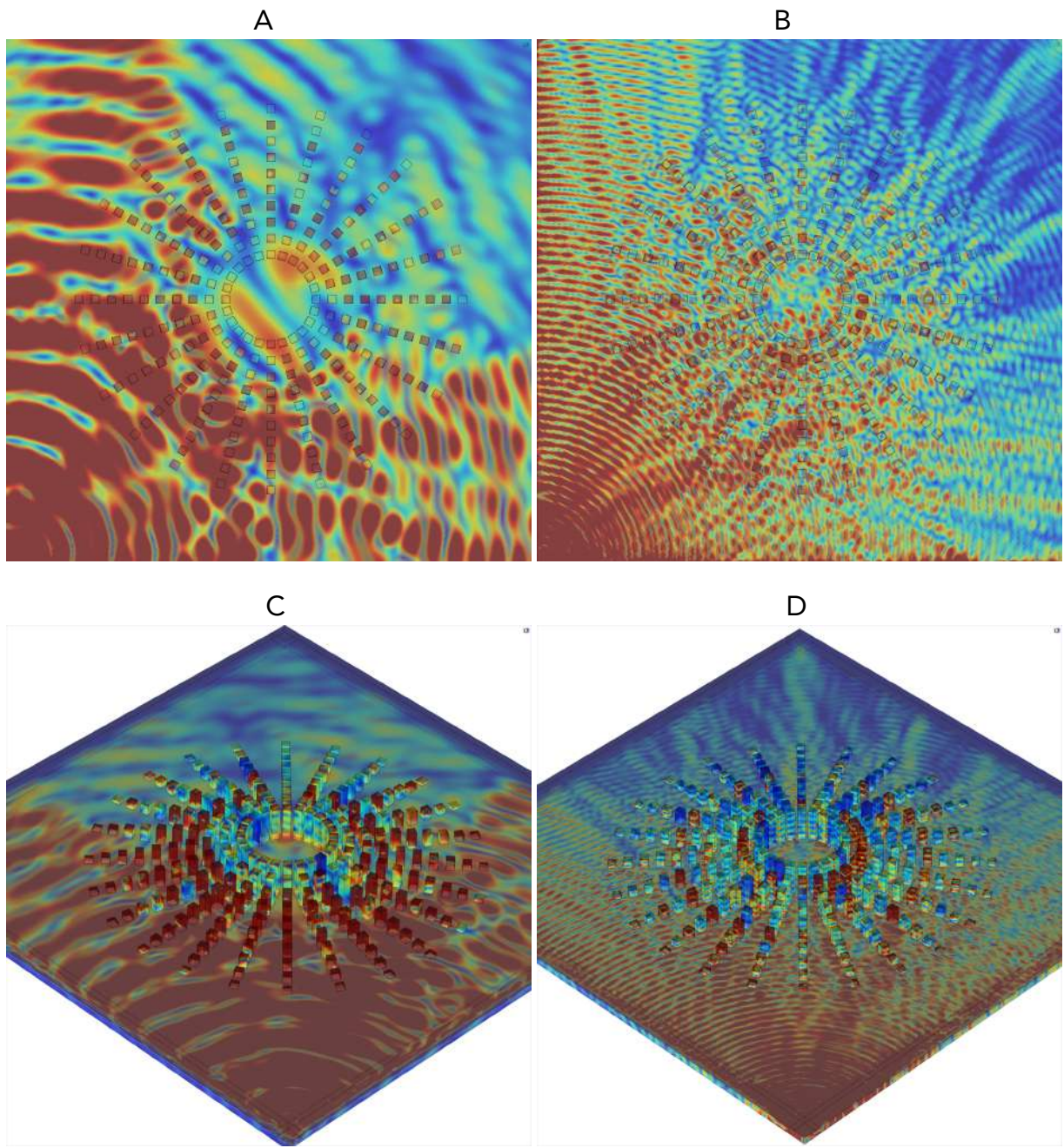

Fig. 4. Scattering of the metacity cloak by a point source at frequency 10 Hz (A) and at 50 Hz (B), situated at a distance of 700 m from the centre of the cloak where (C) and (D) are the 3D views of (A) and (B), respectively. These are concrete buildings on a concrete plate. Color scale is linear and corresponds to total magnitude of total displacement field (from dark blue for vanishing field to dark red for its maximum).

We have performed COMSOL computations for two types of cloaks: The first type consists of concrete buildings on a concrete ($E = 25e9$ [Pa], $\rho = 2300$ [kg/m$^3$], $v = 0.2$) plate, figure 4 and (left panels in Figs. 5-8) and the second type consists of steel ($E = 200e9$ [Pa], $\rho = 7870$ [kg/m$^3$], $v = 0.29$) buildings in a concrete plate (right panels in Figs. 5-8). In figure 4, we start with the most challenging case of a seismic source placed in the intense near-field region of a potential cloak. Clearly, the wavefront behind the cloak is disturbed, but the field amplitude inside the centre of the cloak is clearly reduced. Such an effect is similar to the shielding experiment in [1], although in the present case the shield is an annulus rather than a slab. We shall see later that when the source is placed in the inside of the cloak, the wave pattern outside the cloak is nearly isotropic which is a hallmark of invisibility. So, the situation is more complex than it seems. Let us now consider a seismic source far away from the cloak, it is now a plane wave incident upon the cloak as in figures 5-8. One can see that while almost perfect invisibility is achieved for the first type of cloak (concrete resonators) at 10 Hz (the seismic field is nearly unperturbed outside the cloak, figure 5(A)), the seismic wave field inside its core has a wave pattern reminiscent of what one would expect for a concentrator (the distance between wavefronts is decreased inside the core). However, the second type of cloak (steel resonators) demonstrates good protection features in figure 5(B), but is as neutral as the cloak in (A) for its surroundings. When we increase the frequency, one can see in figure 6 that cloaking is preserved to certain extent at 20 Hz, and the concentrating effect in (A) disappears, whereas the protection in (B) is maintained. However, at 40 Hz, one can see in figure 7 that both cloaks are no longer neutral for their environment. Actually, the second type of cloak in figure 7(B) acts as a seismic wave shield and therefore buildings in front of it would be even more damaged, by the reflection, than in the absence of the cloak, but buildings placed after the cloak are in a shadow, relatively safe, zone, see figure 7(C) which provides another perspective on the wave phenomenon.

These findings are a follow up of the proposal in [4] for shielding a place against seismic waves, by appropriately designed buildings of a city, used as above-ground resonators.

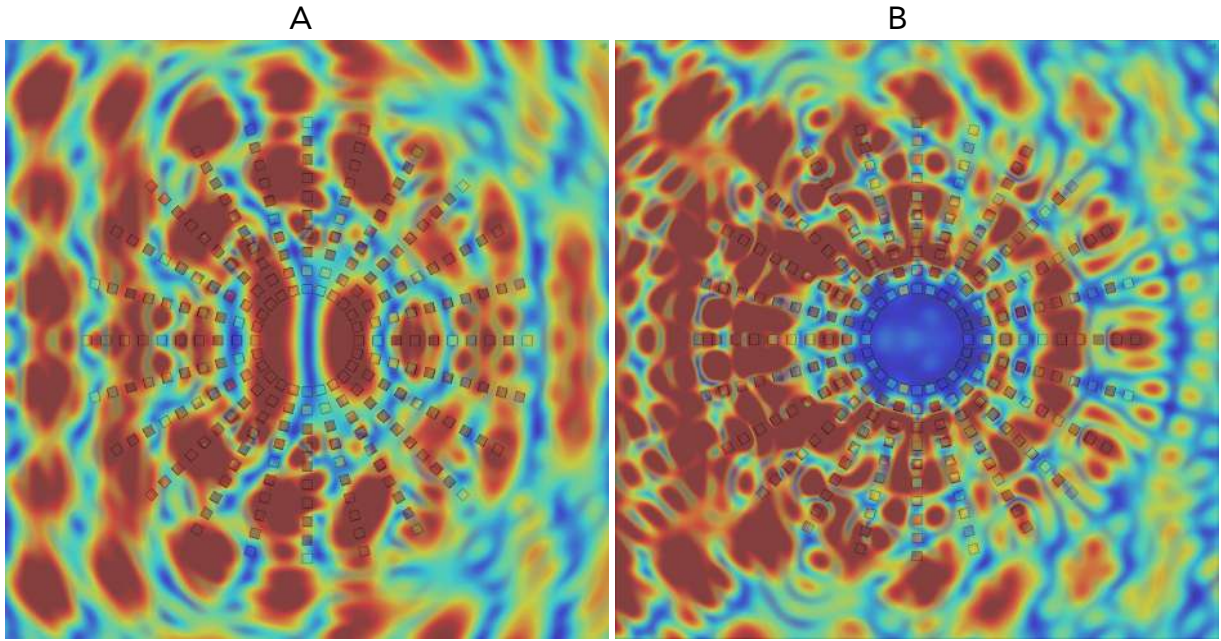

Fig. 5. Scattering of the metacity cloak by a plane wave source at 10 Hz; (A) concrete buildings on a concrete plate (E = 25e9 [Pa], ρ = 2300 [kg/m3], ν = 0.2); (B) steel buildings (E = 200e9 [Pa], ρ = 7870 [kg/m3], ν = 0.29) on a concrete plate. Color scale is as in Figure 4. One notes the antagonistic behaviour of the field inside the core of the cloaks in (A) and (B).

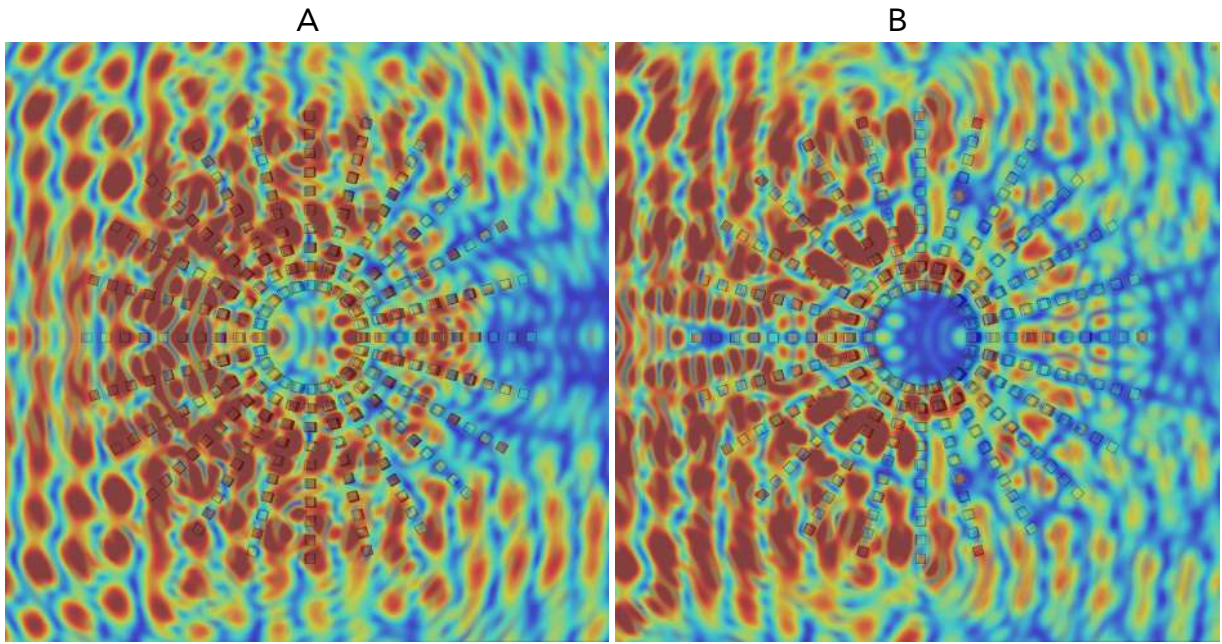

Fig. 6. Scattering of the metacity cloak by a plane wave source at 20 Hz; (A) concrete buildings on a concrete plate (E = 25e9 [Pa], ρ = 2300 [kg/m3], ν = 0.2); (B) steel buildings (E = 200e9 [Pa], ρ = 7870 [kg/m3], ν = 0.29) on a concrete plate. Color scale is as in Figure 5.

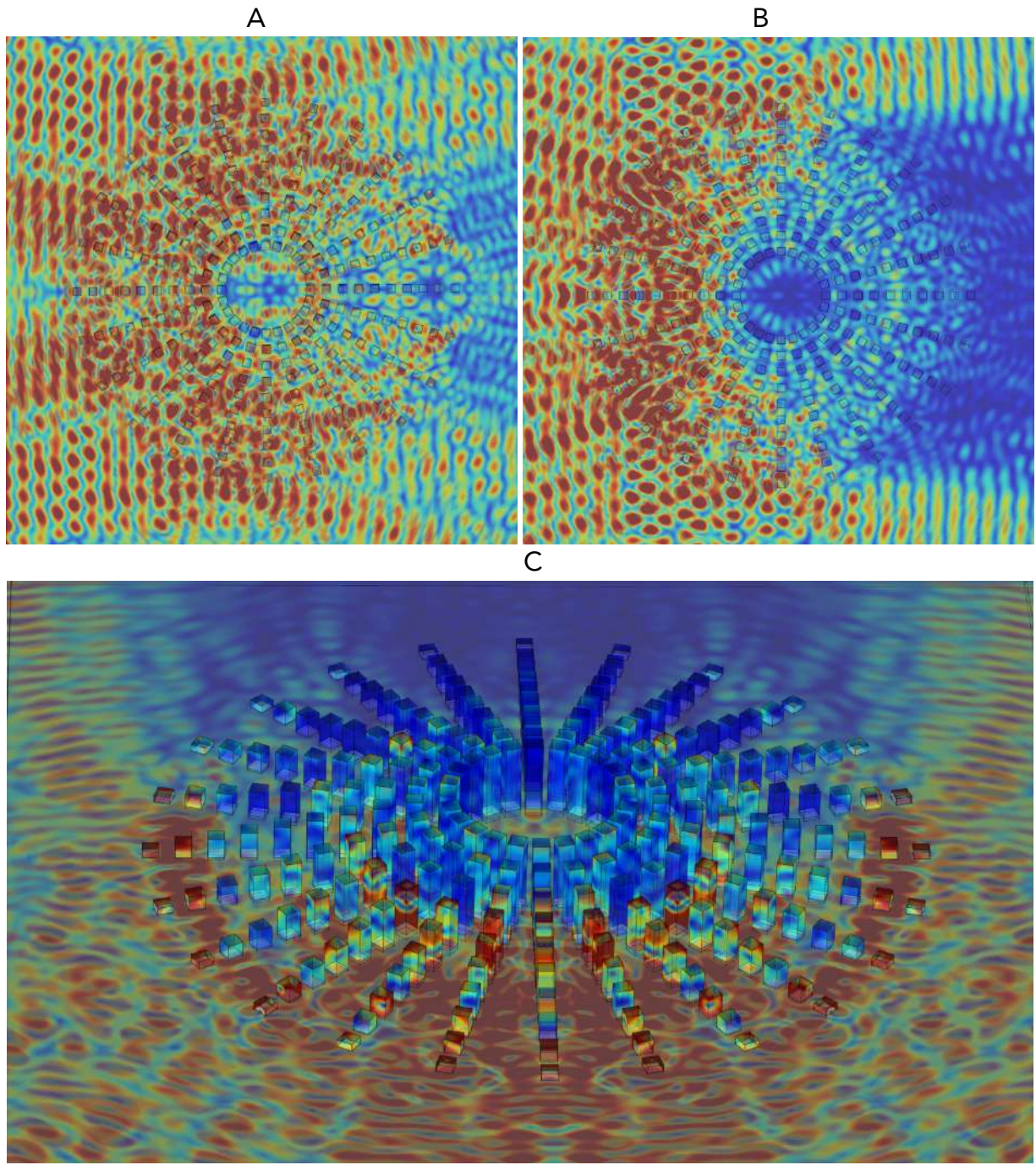

Fig. 7. Scattering of the metacity cloak by a plane wave source at 40 Hz with different materials in (A) and (B): (A) concrete buildings on a concrete plate (E = 25e9 [Pa], ρ = 2300 [kg/m3], ν = 0.2); (B) consists of steel (E = 200e9 [Pa], ρ = 7870 [kg/m3], ν = 0.29) buildings on a concrete plate (right panel). Lower panel is a 3D view of (B).

To show that the first type of cloak (concrete resonators) has indeed good invisibility features, we make a numerical experiment of a source placed inside its core at 10 Hz, see figure 8 (A,C) that shows that the seismic wavefield outside the cloak is almost isotropic, which is a hallmark of cloaking [21]. Obviously, when the frequency gets higher, see figure 8 (B,D) for a source at 50 Hz, homogenization breaks down and the seismic wavefield starts sensing the structural elements within the cloak, which now behaves as many secondary sources and a complex wave pattern outside the cloak with ray-like wave trajectories in radial directions, ensues.

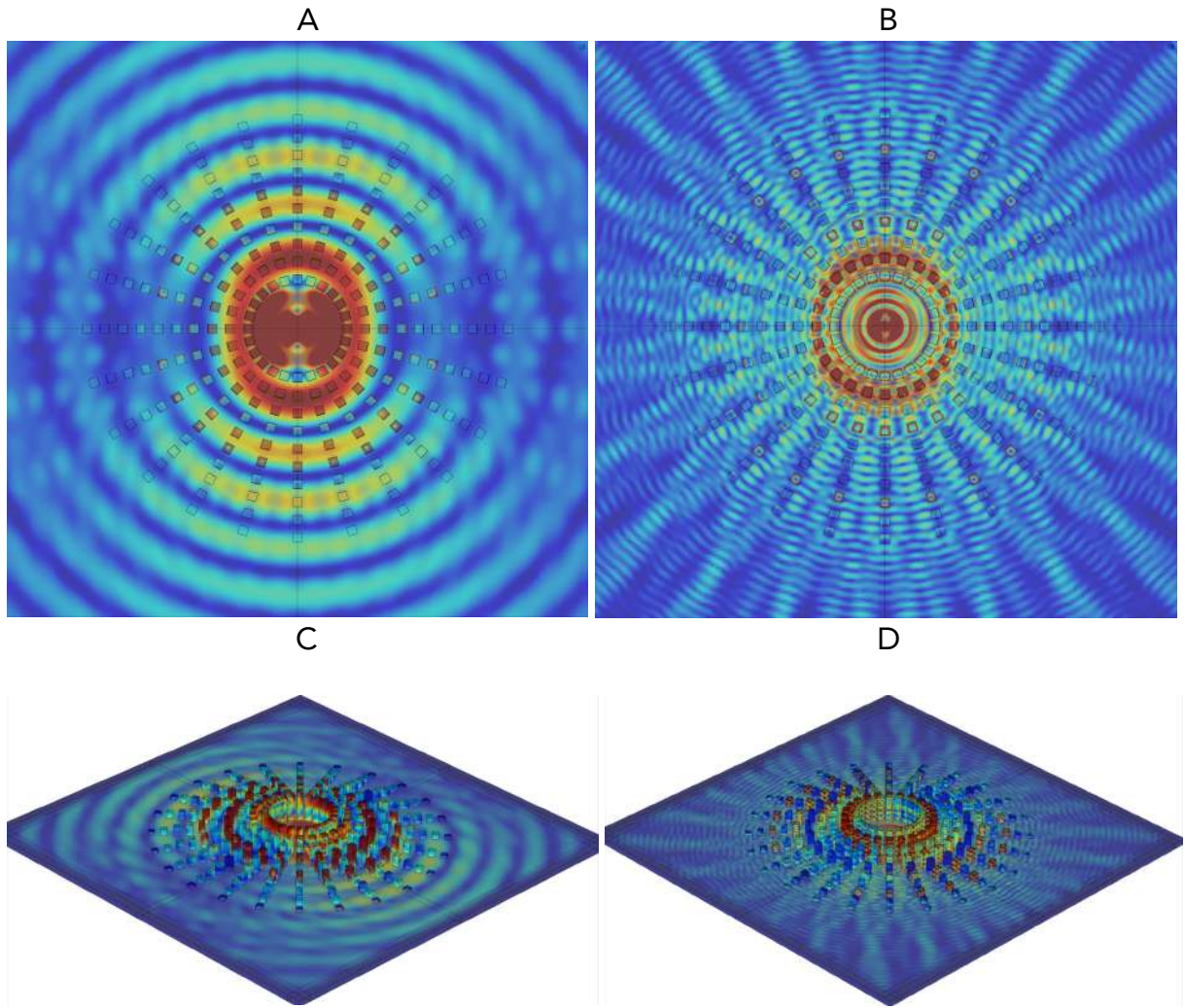

Fig. 8. Cloaking of a seismic source located inside the core of the meta-city cloak at 10 Hz (A,C) and 50 Hz (B,D). These are concrete buildings on a concrete plate. One notices in (A) that the field outside the cloak has reduced amplitude and is almost isotropic, which is a hallmark of cloaking a sensor in optics, while the field in (B) displays some interesting features reminiscent of plasmonic cloaks (buildings located on the outer rings seem to behave as secondary sources).

## 3. Open questions, new ideas, and future trends

We have presented some intriguing results that show buildings do interact with each other and the wavefield and can, under some circumstances, cloak a large scale elastic metamaterial. On the one hand, a perfect cloak would not prevent boundary measurements to be made for complete reconstruction of the elastic properties of a stiff object. In fact, depending upon the type of elastic cloaking theory-- preserving the symmetries of the elasticity tensor at the cost of dealing with a transformed equation of the Willis type [15], or preserving the structure of the elasticity equation at the cost of breaking the symmetries of the elasticity tensor [20]-- leads to different theoretical bounds on elastic cloaking [4]. This could have important consequences in the effectiveness of a seismic protection as the design of seismic cloaks relies upon some structured media that should approximate to certain extent the transformed parameters [11,25-27]. We believe that our two cloaks' (either concrete or steel resonators) can be further improved by better understanding how one can homogenize elastic plates with resonant buildings [38,39]. This is a delicate task which could be achieved by some reverse engineering algorithm using the transformed parameters in Willis's equation (which has fully symmetric rank 4, 3 and 2) tensors and doing same for the Cosserat case.

Our numerical simulations have shown that Young's modulus, Poisson's ratio and density play a prominent role in the way the seismic wave couples with the elements of the cloak. When the plate and its structural elements have same elastic parameters, we achieve good cloaking quality at low frequencies (10 Hz and 20 Hz), but protection is rather poor (at 10 Hz, we just achieve the opposite effect, of a field concentration inside the centre of the cloak). When the plate and its structural elements have markedly different elastic parameters, we achieve poor cloaking quality but good protection. Thus, changing the Young modulus leads to a shielding effect instead of cloaking as it can be seen in figure 7(C), which is very common in reality, buildings having very different properties compared with the alluvial soils, propagation medium of the seismic waves.

In fact, civil engineers compute the eigenmodes of each structure they wish to erect in order to avoid the interaction of the seismic wave with the structure they want to protect. This is achieved by choosing the characteristics of these structures in a way that they will absorb the minimum of the energy transported by the elastic wave so as to avoid the resonance.

Our proposal is that we should simply do the opposite to build structures in a smart way, in such a way that they will absorb exactly what we need to change the properties of these surface Rayleigh waves and so to obtain zones with zero seismic impact - the cloaked zones.

Another possibility is to convert and redirect the surface Rayleigh waves thanks to the transformation elasticity theory, or by filtering the surface waves frequencies thanks to the elastic rainbow effect [5].

Literature in geophysics and earthquake engineering is vast but ideas supporting the importance of site-city interaction [3, 10, 40-42] are starting to be taken into consideration. It will take more time to convince the earthquake engineering community that ideas borrowed from the metamaterials community, notably regarding invibility and cloaking [43], can help in the design of new cities.

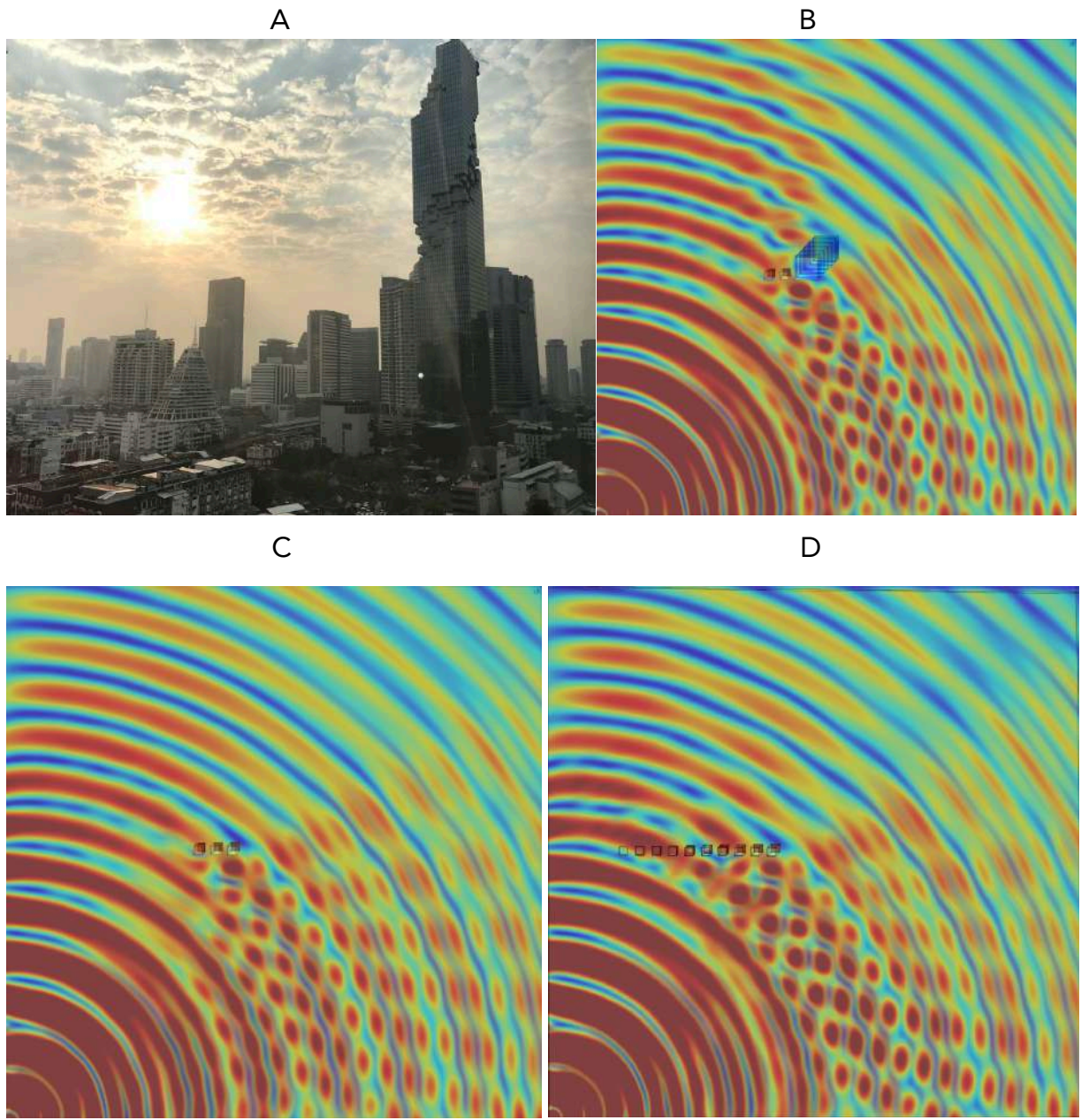

Fig. 9. Photo of a skyscraper in Bangkok (A) which stands out (courtesy of B. Ungureanu). This building, which height is 300 m, of square cross section 60 m x 60 m, should interact with an incoming seismic wave at 10 Hz according to a COMSOL simulation (B).

(C) and (D) simulations shows the difference of the propagation of surface R waves considering 3 identical structures or 10 structures with height varying linearly from 10 m to 100 m, of square cross section 20 m x 20 m, with centre to centre spacing of 40m along the propagation path, which behave like a wave guide.

In conclusion, the advantages of cloaking versus that of shielding protection is a non-trivial topic, which has hardly been addressed in the existing literature. The field of seismic metamaterials is still in its infancy and further theoretical and experimental results are required to answer the level of seismic protection one could achieve with structured soils in civil engineering. One could, for instance, envision that a skyscraper like in figure 9 may act as a directive antenna when an earthquake hits the megalopolis. This skyscraper might not only vibrate tremendously, but it might also act as a secondary seismic source which would modify the dynamic response of peripheral buildings. The way that architects could design cities of the future, with higher and higher buildings, might actually have deep consequences in terms of local seismicity and this study, and other previous ones based on geophysical measurements in situ, shows that one might wish to design a structure taking into account its close neighbours.

ACKNOWLEDGEMENT

Richard V. Craster acknowledges support of CNRS through the Unité Mixte Internationale Abraham de Moivre. Bogdan Ungureanu acknowledges funding of European Union (MARIE SKŁODOWSKA-CURIE ACTIONS project Acronym/Full Title: METAQUAKENG - Metamaterials in Earthquake Engineering - MSCA IF - H2020).